\documentclass[fleqn,usenatbib]{mnras}

\usepackage{newtxtext,newtxmath}

\usepackage[T1]{fontenc}

\DeclareRobustCommand{\VAN}[3]{#2}
\let\VANthebibliography\thebibliography
\def\thebibliography{\DeclareRobustCommand{\VAN}[3]{##3}\VANthebibliography}

\usepackage{graphicx}	
\usepackage{amsmath}	
\usepackage[UKenglish]{babel}
\DeclareRobustCommand{\HC}{\ensuremath{\mathrm{HC}_5\mathrm{NH}^+}}
\DeclareRobustCommand{\HH}{\ensuremath{\mathrm{H}_2}}

\title[Rotational Excitation of HC$_5$NH$^+$ by He]{Rotational Excitation of HC$_5$NH$^+$ by He: Collision Dynamics and Astrophysical Applications}

\author[Xiang et al.]{
	Ziyu Xiang$^{1}$, Liman Tian$^{1}$, Tong Sun$^{1}$, Chuangang Ning$^{2}$
	and Xiaolin Chen$^{1}$\thanks{E-mail: chenxl13@zafu.edu.cn}
	\\
	$^{1}$College of Optical, Mechanical and Electrical Engineering (College of Intelligent Equipment Engineering), Zhejiang A \& F University, Hangzhou ,311300, PR China\\
	$^{2}$Department of Physics, Tsinghua University, Beijing, 100084, PR China\\
}

\date{Accepted XXX. Received YYY; in original form ZZZ}
\pubyear{2026}

\begin{document}
	\label{firstpage}
	\pagerange{\pageref{firstpage}--\pageref{lastpage}}
	\maketitle
	
	\label{firstpage}
	\pagerange{\pageref{firstpage}--\pageref{lastpage}}
	\maketitle
	\thispagestyle{empty}

	\begin{abstract}
	The protonated cyanopolyyne HC$_5$NH$^+$ is an interstellar molecular ion of astrochemical interest, and understanding its rotational excitation is essential for interpreting its emission in molecular clouds. In this work, we construct a two-dimensional potential energy surface (PES) for the HC$_5$NH$^+$--He system using the CCSD(T)-F12B/cc-pVTZ-F12 method. Rotationally inelastic cross sections are calculated using the quantum close-coupling (CC) method, and the corresponding collisional rate coefficients are obtained by thermal averaging over a Maxwell--Boltzmann distribution. These rate coefficients are incorporated into a non-local thermodynamic equilibrium (non-LTE) radiative transfer model to investigate the excitation of HC$_5$NH$^+$ in the cold dark cloud TMC-1. A global $\chi^2$ minimization analysis of seven observed transitions yields best-fitting values of $n(\mathrm{H}_2)=1.63\times10^{5}$ cm$^{-3}$ and $N(\mathrm{HC}_5\mathrm{NH}^+)=5.09\times10^{11}$ cm$^{-2}$, with a minimum reduced $\chi^2$ of 0.87. The modelled line intensities are in good agreement with the observations. This study provides the first-principles collisional data needed to model the rotational excitation of HC$_5$NH$^+$ and offers new constraints on its physical and spectroscopic properties in TMC-1.
	\end{abstract}
	
	
	\section{Introduction}
	\label{sec:intro}
	
	Protonation gives molecules large permanent dipole moments and small rotational constants, making the resulting ions detectable radio tracers in cold dark clouds and providing indirect evidence for their neutral parents \citep{agundez2013}. Among the carbon-chain species found in these environments, cyanopolyynes and their protonated counterparts, HC$_{2n+1}$N and HC$_{2n+1}$NH$^{+}$ ($n=0,1,2,\ldots$), form an important homologous series. While neutral cyanopolyynes have been extensively studied \citep{mcguire2018}, their protonated counterparts have been far less extensively detected. To date, the members reported in the protonated series include HCNH$^{+}$ and the longer-chain species with $n=1$--3 \citep{ziurys1986,kawaguchi1994,marcelino2020,cabezas2022}, with their column densities generally decreasing as the carbon-chain length increases.
	
	\HC\ is the focus of this work. Its proposed interstellar detection is based on seven lines with nearly harmonic frequency spacing found in the Yebes 40~m survey of TMC-1 \citep{marcelino2020}, with \HC\ as their most likely carrier. This assignment is supported by theoretical rotational constants and by the similarity between the abundance ratios of HC$_5$N/\HC\ and HC$_3$N/HC$_3$NH$^{+}$. However, in the absence of laboratory rotational spectroscopy, its detection remains tentative. Laboratory infrared spectroscopy has provided direct frequency references for the identification of HC$_3$NH$^{+}$ \citep{lee1987,kawaguchi1990}, whereas the assignments of longer-chain protonated species have relied largely on rotational constants and dipole moments predicted by high-level coupled-cluster calculations \citep{botschwina1997,heyl1997}. We therefore adopt this assignment as a working hypothesis for the collisional and radiative transfer analyses.
	
	The observationally inferred column densities of these ions depend on the rotational level populations, and collisional excitation in cold dark clouds is generally insufficient to maintain local thermodynamic equilibrium (LTE). Studies of HCNH$^{+}$ collisions with He and \HH, together with hyperfine-resolved data for its isotopologues, have demonstrated how dedicated potential energy surfaces (PESs) and state-to-state rate coefficients affect spectral predictions \citep{bop2023,bop2024,bop2026}. However, because accurate rate coefficients for long-chain molecules are extremely demanding to compute, quantum scattering results have so far been reported only for HC$_3$NH$^{+}$--He \citep{almogren2021}, and none are available for \HC\ or longer chains. Moreover, the effects of increasing chain length on the potential anisotropy and transition propensities cannot be directly extrapolated from shorter-chain systems.
	
	In this work, we construct a two-dimensional \textit{ab initio} PES for \HC--He and perform CC calculations to obtain rotationally inelastic cross sections and low-temperature rate coefficients. The resulting collisional data are then used in non-LTE radiative transfer calculations and in fitting the observed line intensities towards TMC-1, in order to determine its excitation properties and constrain its column density. The remainder of this paper is organised as follows. Section~\ref{sec:methods} describes the construction and Legendre expansion of the \HC--He PES. Section~\ref{sec:scattering} presents the calculated inelastic cross sections and rate coefficients. Section~\ref{sec:astro} discusses the astrophysical applications and the non-LTE radiative transfer results, and Section~\ref{sec:conclusions} summarises our conclusions.
	
	\section{Theoretical methods and computational details}
	\label{sec:methods}
	
	\subsection{Potential energy surface}
	\label{sec:pes}
	
	Within the rigid-rotor approximation, the molecular geometry of \HC\ is fixed at the equilibrium structure of its ground state, and only its overall rotation is considered. The lowest vibrational frequency of the ion is approximately 99\,cm$^{-1}$ \citep{marcelino2020}. At a typical interstellar temperature of 20\,K, the Boltzmann factor for the first vibrationally excited state, $\exp(-99/k_{\mathrm{B}}T)$, is only about 0.1\%. Vibrational excitation can therefore be neglected over the temperature range considered in this work \citep{lique2006,faure2020}. This approximation has been widely adopted for similar weakly interacting systems, including SiC$_4$--He \citep{chefai2024}, HNSi/HSiN--He \citep{tian2026a}, and CaO/CaS--He \citep{tian2026b}.
	
\begin{figure}
		\centering
		\includegraphics[width=\columnwidth]{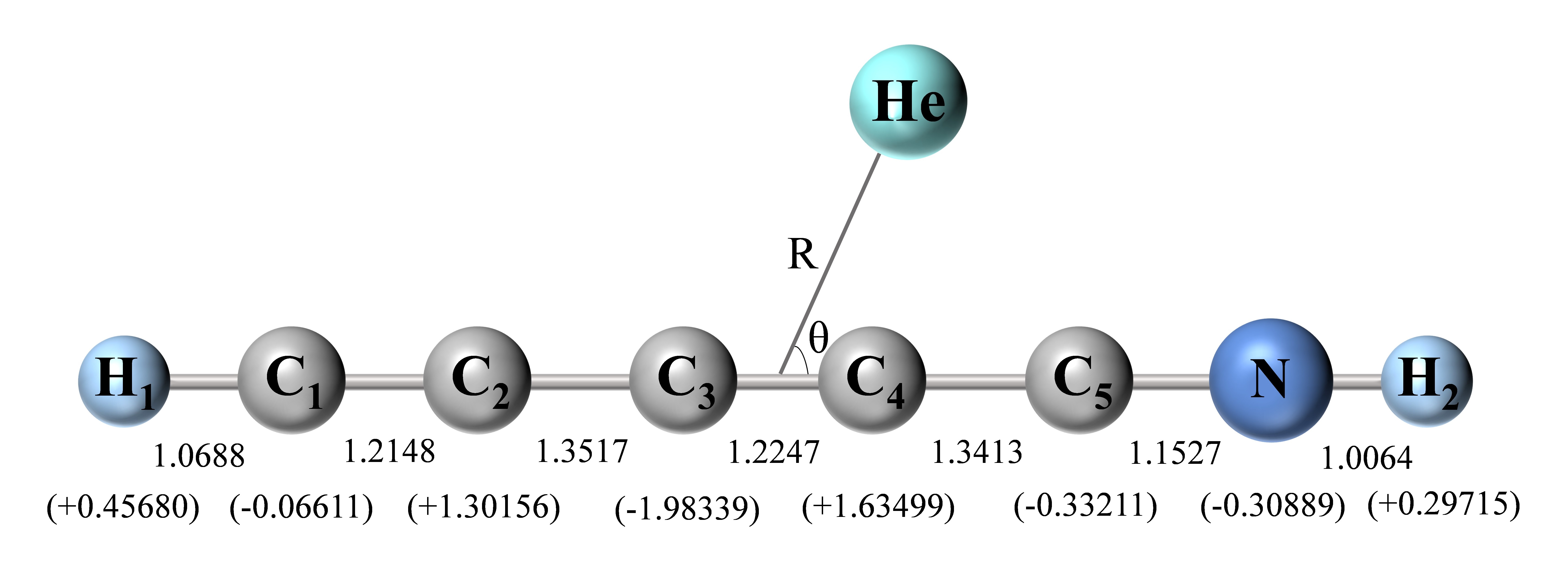}
		\caption{Geometry of the \HC--He collision system and the Jacobi coordinates.}
		\label{fig:geometry}
	\end{figure}

	\HC\ is a linear conjugated polyyne ion. In the centre-of-mass frame, the position of He relative to the linear \HC\ molecule in its electronic ground state is described by the Jacobi coordinates ($R,\theta$), where $R$ is the distance between He and the centre of mass of the ion, and $\theta=0^\circ$ corresponds to the arrangement  HC$_5$NH$^{+}$ ---He, as illustrated in Fig.~\ref{fig:geometry}. The bond lengths are calculated at the CCSD(T)-F12B/cc-pVTZ-F12 level of theory and are as follows:	
	$r_{\text{H$_1$C$_1$}}=1.0688$\,\AA, $r_{\text{C$_1$C$_2$}}=1.2148$\,\AA, 
	$r_{\text{C$_2$C$_3$}}=1.3517$\,\AA,
	$r_{\text{C$_3$C$_4$}}=1.2247$\,\AA,  $r_{\text{C$_4$C$_5$}}=1.3413$\,\AA, 
	 $r_{\text{C$_5$N}}=1.1527$\,\AA, and  $r_{\text{NH$_2$}}=1.0064$\,\AA.
	
	The interaction energy is calculated using the explicitly correlated coupled-cluster method CCSD(T)-F12B and the cc-pVTZ-F12 basis set \citep{knizia2009,werner2012}. The radial grid covers $R=2$--30\,\AA, spanning the short-, intermediate-, and long-range regions. The angle $\theta$ is sampled uniformly from $0^\circ$ to $180^\circ$ in steps of $5^\circ$, yielding a total of 2553 geometries. The basis-set superposition error (BSSE) is corrected using the Boys--Bernardi counterpoise scheme:
	\begin{equation}
		V(R,\theta)=E_{\mathrm{HC_5NH^+}\text{--}\mathrm{He}}(R,\theta)
		-E_{\mathrm{HC_5NH^+}}(R,\theta)-E_{\mathrm{He}}(R,\theta),
		\label{eq:interaction}
	\end{equation}
	where $E_{\mathrm{HC_5NH^+}\text{--}\mathrm{He}}$, $E_{\mathrm{HC_5NH^+}}$, and $E_{\mathrm{He}}$ are the energies of the complex, the \HC\ monomer, and the He atom, respectively, all calculated within the same set.
	
\begin{figure}
		\centering
		\includegraphics[width=\columnwidth]{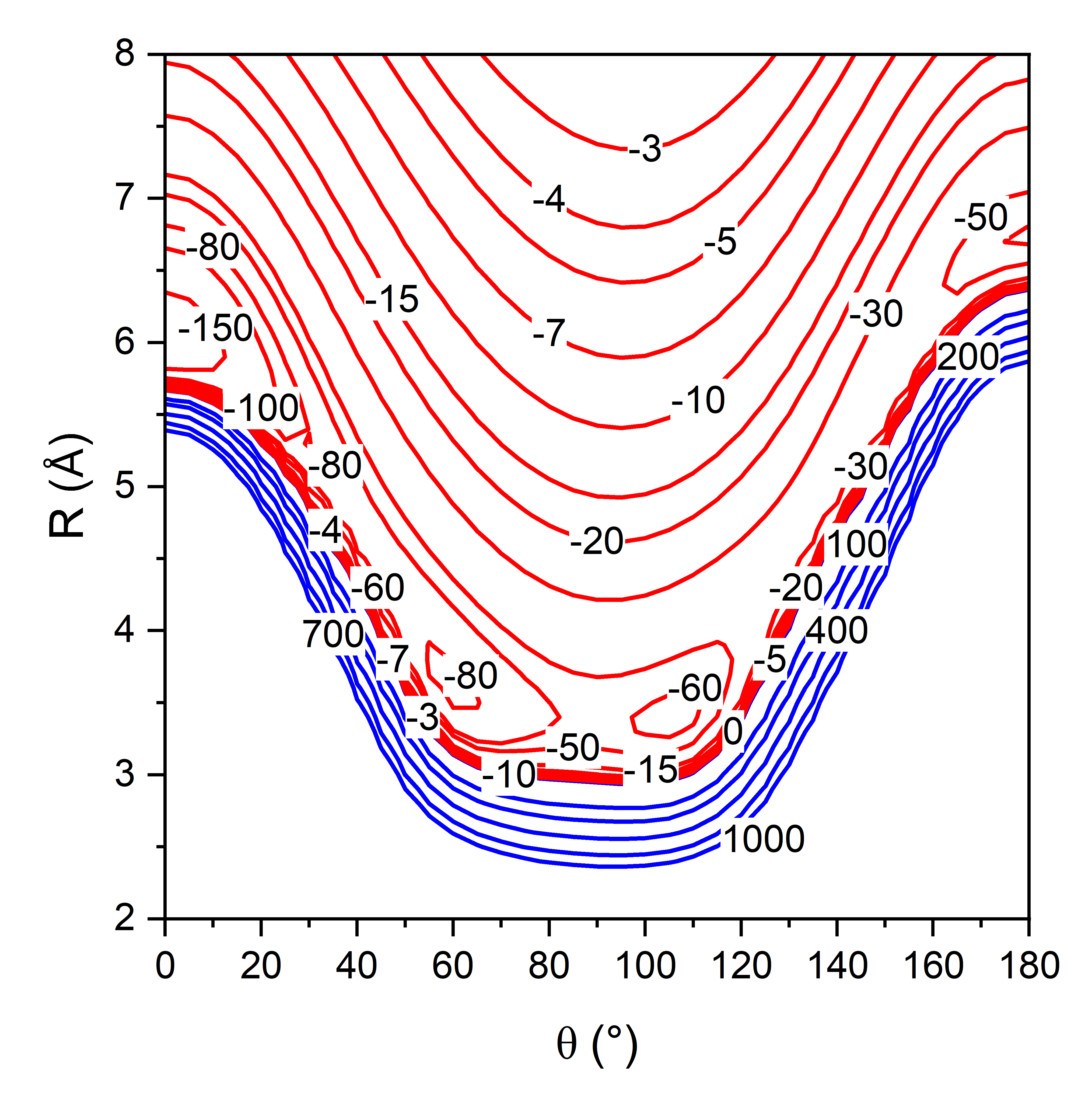}
		\caption{Contour plot of the \HC--He interaction potential $V(R,\theta)$ (in cm$^{-1}$). Blue contours indicate positive values, while red contours indicate negative values.}
		\label{fig:pes}
	\end{figure}

	The resulting PES is shown in Fig.~\ref{fig:pes}. It exhibits pronounced axial asymmetry and a lateral double-minimum structure. Its global minimum lies on the axis at the protonated end ($\theta=0.2^\circ$), reaching a depth of 194.55\,cm$^{-1}$ at $R=6.0$\,\AA. By contrast, the well at the N end ($\theta=180^\circ$) is much shallower. Laterally, the interaction exhibits two asymmetric minima, separated by a shallow saddle point near $\theta=90^\circ$ with a depth of about 58\,cm$^{-1}$; consequently, the 60\,cm$^{-1}$ contour in the lateral region splits into two closed loops. The PES exhibits four distinct low-energy regions associated with the H$_1$ (+0.45680), C$_2$ (+1.30156), C$_4$ (+1.63499), and H$_2$ (+0.29715) sites of \HC, corresponding to the four main low-energy regions in the contour map. Their spatial distribution reflects the pronounced anisotropy and end-group asymmetry of the interaction potential, with the deepest well located near the protonated H$_2$ end.
	
	\subsection{Legendre expansion of the PES}
	\label{sec:legendre}
	
	In the short-range region, the raw interaction energy $V_a$ exhibits an unphysically steep increase due to orbital overlap and the numerical behaviour of the Boys--Bernardi correction. To address this issue, before performing the Legendre expansion, we apply the smoothing and scaling method of \citet{gonzalez2024} to the raw \textit{ab initio} energies. Denoting the raw and smoothed energies by $V_a$ and $V_s$, respectively, the procedure is defined as
	\begin{equation}
		V_s=
		\begin{cases}
			V_a, & V_a\le V_{T1},\\[4pt]
			V_{T1}+(V_{T2}-V_{T1})\,
			S\!\left(\dfrac{V_a-V_{T1}}{V_{T2}-V_{T1}}\right),
			& V_{T1}<V_a\le V_{T2},\\[6pt]
			V_{T1}+\dfrac{2}{\pi}(V_{T2}-V_{T1}), & V_a>V_{T2},
		\end{cases}
		\label{eq:smoothing}
	\end{equation}
	where the threshold parameters are $V_{T1}=2000$\,cm$^{-1}$ and $V_{T2}=10\,000$\,cm$^{-1}$, and the switching function is \(S(x) = \frac{2}{\pi} \sin\!\left[ \frac{\pi}{2} \sin\!\left( \frac{\pi}{2} x \right) \right]\).
	This scheme ensures that $V_s$ and its first derivative are continuous at the joining points, while limiting the high-energy repulsive wall to a finite value, thereby avoiding numerical instabilities in the scattering calculations.
	
	For use in the scattering calculations, the two-dimensional PES is expanded in Legendre polynomials:
	\begin{equation}
		V(R,\theta)=\sum_{\lambda=0}^{\lambda_{\max}}
		V_\lambda(R)\,P_\lambda(\cos\theta),
		\label{eq:legendre}
	\end{equation}
	where $P_\lambda$ is the Legendre polynomial of order $\lambda$ and $V_\lambda(R)$ are the radial coefficients. To reduce the computational cost while maintaining accuracy, the expansion is truncated at $\lambda_{\max}=56$, for which the mean deviation of the reconstructed PES from the \textit{ab initio} values is less than 0.5\%.
	
\begin{figure*}
		\centering
		\includegraphics[width=\textwidth]{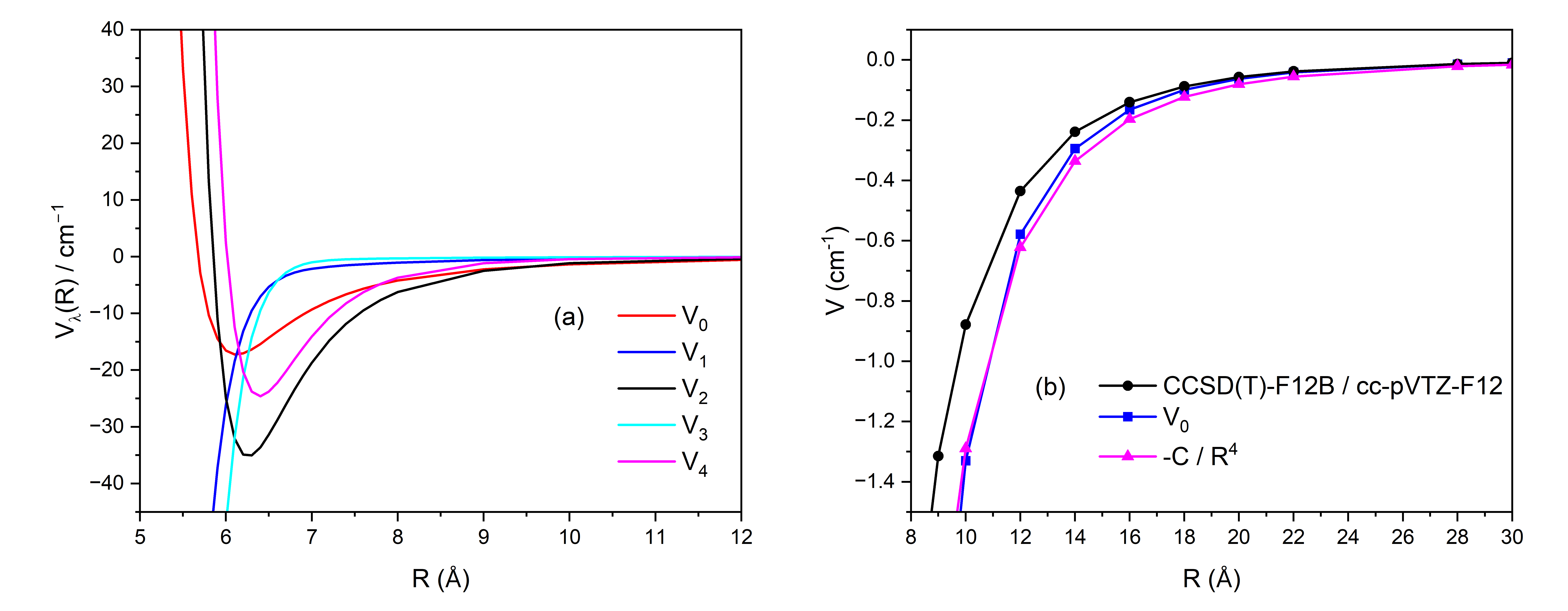}
		\caption{(a) Radial dependence of the first five Legendre radial coefficients $V_\lambda(R)$ ($\lambda=0$--4, in cm$^{-1}$), as functions of $R$ for the \HC\ system. (b) Radial dependence of the long-range interaction potential.}
		\label{fig:radial}
	\end{figure*}

	Fig.~\ref{fig:radial}(a) shows the $R$ dependence of the first five radial coefficients. Among the anisotropic terms with $\lambda>0$, $V_2$ has a markedly larger amplitude than the others, followed by $V_4$, indicating the dominance of the even-order terms. The odd terms $V_1$ and $V_3$ have appreciable amplitudes only at short range and decay rapidly with increasing $R$. The dominance of the even-order terms reflects the strong long-range induction and dispersion attraction experienced by He as it approaches the conjugated carbon chain from the side, which also explains the propensity for $\Delta j=\pm2$ transitions during collisions.
	
	The long-range fit provides a further test of the asymptotic behaviour of the PES. Fig.~\ref{fig:radial}(b) shows that, in the long-range region $R\ge16$\,\AA, the \textit{ab initio} points agree well with both $V_0$ and the $-C/R^4$ fit, consistent with an $R^{-4}$ asymptote dominated by the charge--induced-dipole interaction \citep{buckingham1978}. For $R<16$\,\AA, the \textit{ab initio} points lie systematically above the $-C/R^4$ asymptote, indicating that contributions from repulsion and higher-order terms such as $R^{-6}$ are no longer negligible. This also explains why the analytic potential adopts the asymptotic form only in the long-range region, while the inner region is described by a Legendre expansion. The two branches are joined such that their first derivatives are continuous, avoiding unphysical kinks or barriers in the low-energy scattering calculations. The deviation between the \textit{ab initio} points and the fit decreases rapidly with increasing $R$ and tends to zero, supporting the use of this analytic PES in subsequent low-energy scattering calculations.
	
	\section{Scattering cross sections and rate coefficients}
	\label{sec:scattering}
	
	Using the PES described above, we employ the MOLSCAT code \citep{hutson2019} to calculate the inelastic scattering of an atom with \HC\ within the CC framework \citep{arthurs1960}, obtaining state-to-state cross sections for collision energies up to 300\,cm$^{-1}$. For \HC, we adopt the rotational constant $B_0=0.043224$\,cm$^{-1}$ and the centrifugal distortion constant $D_0=9.11\times10^{-10}$\,cm$^{-1}$ \citep{marcelino2020}.
	
	The partial-wave sum for the integral cross section from $j$ to $j'$ is given by:
	\begin{equation}
		\begin{split}
			\sigma_{j\to j'}(E)={}&
			\frac{\pi}{k_j^2(2j+1)}
			\sum_{J=0}^{\infty}
			\sum_{l=|J-j|}^{J+j}
			\sum_{l'=|J-j'|}^{J+j'}(2J+1)\\
			&\times
			\left|
			\delta_{jj'}\delta_{ll'}
			-S^{J}_{jj',\,ll'}(E)
			\right|^2 ,
		\end{split}
		\label{eq:cross}
	\end{equation}
	where $E$ is the collision energy, $J$ is the total angular momentum, $l$ and $l'$ are the orbital angular momenta of the initial and final channels, respectively, and $\delta$ is the Kronecker delta. The channel wavenumber is defined as $k_j=[2\mu(E-E_j)]^{1/2}/\hbar$, where $\mu=3.8023948$\,amu (unified atomic mass units) is the reduced mass of the \HC--He system. The rotational energy levels $E_j$ of the molecule are given by:
	\begin{equation}
		E_j = B_0 j(j+1) - D_0 j^2(j+1)^2
	\end{equation}
	
	To ensure the convergence of transition cross sections involving rotational levels up to \(j \le 20\), the rotational basis truncation parameter \(j_{\max}\) was systematically optimized for different energy ranges. Accordingly, an energy-dependent rotational basis was adopted: \(j_{\max} = 29\) for \(E \le 100\ \mathrm{cm^{-1}}\) and \(j_{\max} = 35\) for \(100 < E \le 300\ \mathrm{cm^{-1}}\). For the transitions considered, the numerical uncertainty of the computed cross sections using these \(j_{\max}\) values is less than 5\%, validating that the adopted basis provides sufficiently converged cross sections for the target levels. The corresponding integration step size control ensures 50, 20, and 10 steps per half-wavelength for \(E = 0\text{--}20\), \(20\text{--}100\), and \(100\text{--}300\ \mathrm{cm^{-1}}\), respectively. To adequately sample the threshold and resonance regions, the collision-energy grid spacings are set to 0.05, 0.1, 1, and \(5\ \mathrm{cm^{-1}}\) for the energy intervals of \(E \le 30\), \(30 < E \le 80\), \(80 < E \le 150\), and \(150 < E \le 300\ \mathrm{cm^{-1}}\), respectively. At each energy point, the total angular momentum \(J_{\mathrm{tot}}\), controlled by JTOTU, is iteratively increased until convergence is reached (with an upper limit set to 99999). The summation process terminates automatically when the maximum changes in both diagonal and off-diagonal cross sections fall below their respective thresholds (\( \mathrm{DTOL} = 0.1\ \mathrm{\mathring{A}^2} \), \( \mathrm{OTOL} = 0.005\ \mathrm{\mathring{A}^2} \)). To fully resolve the resonance features, the radial coupled equations are divided into two propagation stages: a log-derivative propagator is used in the short-range region \citep{manolopoulos1986}, and a variable-step Airy propagator is used in the long-range region \citep{alexander1987}. All calculations are solved using the MOLSCAT program \citep{hutson2019}.
	
\begin{figure*}
		\centering
		\includegraphics[width=\textwidth]{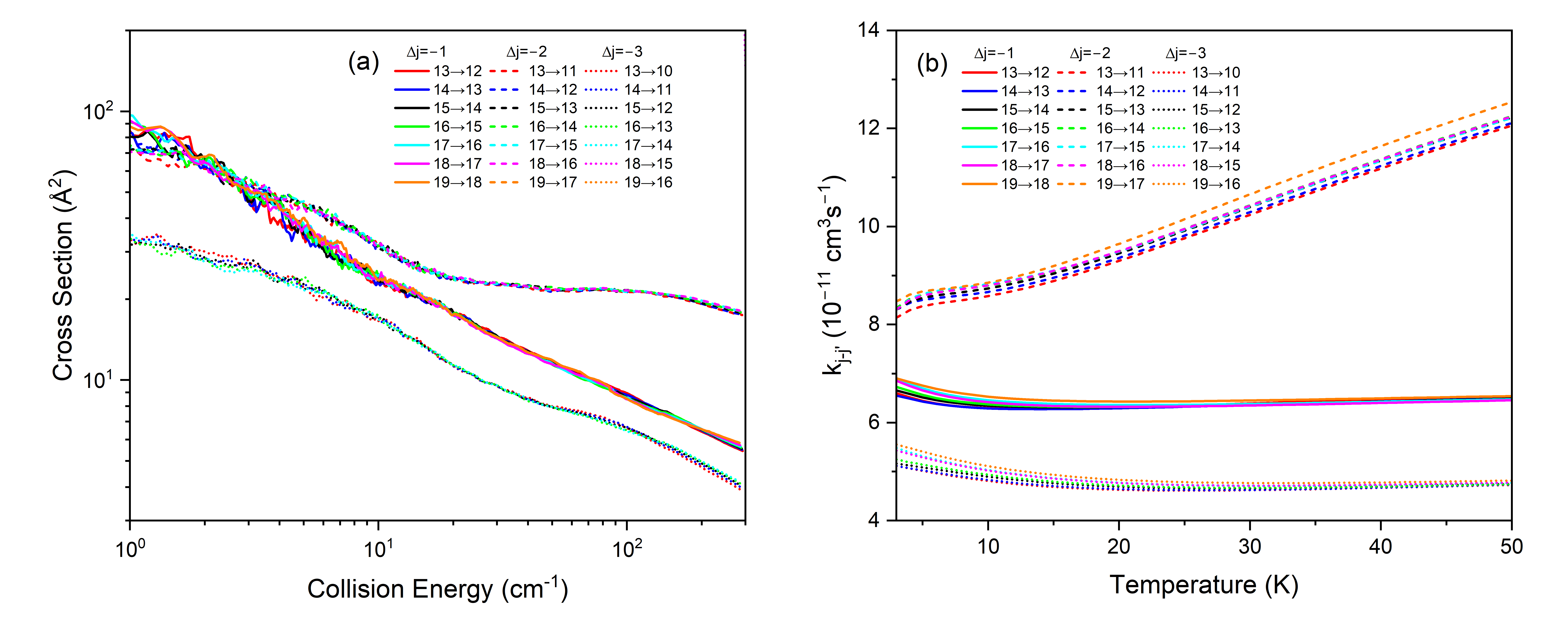}
		\caption{(a) De-excitation integral cross sections for initial states $j=13$--19 as functions of collision energy (on a log--log scale). (b) The corresponding de-excitation rate coefficients as functions of temperature.}
		\label{fig:cross}
	\end{figure*}

	Fig.~\ref{fig:cross} shows the de-excitation dynamics for high-$j$ initial states, and Fig.~\ref{fig:cross}(a) presents the de-excitation cross sections for $j=13$--19. Below 10\,cm$^{-1}$, shape resonances associated with tunnelling through the centrifugal barrier and Feshbach resonances arising from quasi-bound states in the potential wells produce dense oscillations \citep{almogren2021}; at very low energies, the cross section decreases rapidly with increasing energy, approximately following the long-range capture $E^{-1/2}$ scaling \citep{gioumousis1958}. Above 30\,cm$^{-1}$, the cross sections depart from this scaling and decline slowly, indicating that long-range capture is no longer the limiting factor and that the dynamics become governed by the short-range anisotropic potential and exchange repulsion. At the low-energy end, the $\Delta j=-1$ cross section is the largest, whereas the $\Delta j=-3$ cross section is the smallest over most of the energy range; as the energy increases, the decrease in the $\Delta j=-2$ cross section becomes markedly less pronounced and its cross section becomes the largest, with the clearest advantage at high energy. This energy-dependent selectivity is consistent with the anisotropy of the PES being dominated by the $V_2$ term.
	
	Maxwell--Boltzmann averaging of these state-to-state cross sections over a 1\,K grid from 3 to 50\,K yields the state-to-state (de-)excitation rate coefficients for $j\le19$:
	\begin{equation}
		k_{j\to j'}(T)=
		\left(\frac{8}{\pi\mu k_{\mathrm{B}}^3T^3}\right)^{1/2}
		\int_0^\infty
		E_k\,\sigma_{j\to j'}(E_k)\,
		e^{-E_k/k_{\mathrm{B}}T}\,\mathrm{d}E_k ,
	\end{equation}
	where $k_{j\to j'}$ is the de-excitation rate coefficient from the upper level $j$ to the lower level $j'$, and $k_{\mathrm{B}}$ is the Boltzmann constant. The corresponding excitation rate coefficients, $k_{j'\to j}$, are obtained from the principle of detailed balance \citep{benabdallah2012}:
	\begin{equation}
		k_{j'\to j}(T)=
		\frac{2j+1}{2j'+1}
		\exp\!\left(
		-\frac{E_j-E_{j'}}{k_{\mathrm{B}}T}
		\right)
		k_{j\to j'}(T).
		\label{eq:detailedbalance}
	\end{equation}
	
	Fig.~\ref{fig:cross}(b) shows the temperature dependence of the de-excitation rate coefficients: the $\Delta j=-2$ channel increases monotonically and substantially with temperature, becoming dominant at higher temperatures. This behaviour is directly related to its cross section, which decreases most slowly at high collision energies and forms a plateau above 30\,cm$^{-1}$ (Fig.~\ref{fig:cross}a): as the temperature rises, high-energy collisions gain weight and the rate continues to grow. The $\Delta j=-1$ cross section decreases steadily with energy, and its thermally averaged rate coefficient becomes nearly constant above approximately 20\,K, consistent with ion--induced-dipole Langevin capture \citep{gioumousis1958}; the $\Delta j=-3$ channel has the smallest rate and the weakest temperature dependence. Differences among initial states within the same $\Delta j$ are small; only for $\Delta j=-2$ does the rate coefficient increase slightly with increasing initial $j$, following the same trend as the high-energy cross sections in Fig.~\ref{fig:cross}(a). Thermal averaging smooths out the dense resonances at low collision energies, resulting in smooth temperature dependences of the rate coefficients. The weak dependence on the initial rotational level also indicates that this system is less sensitive to the rotational quantum number than neutral molecular systems.
	
	\section{Astrophysical applications}
	\label{sec:astro}
	
	In low-density clouds such as TMC-1, collisional excitation is generally insufficient to fully thermalise the level populations \citep{bop2025}, so the column density derived by \citet{marcelino2020} under the LTE assumption may be subject to systematic bias. Using the rate coefficients derived above, we can quantitatively assess this effect with non-LTE modelling.
	
	Non-LTE radiative transfer simulation is computed with RADEX \citep{vandertak2007}. The code adopts the escape-probability approximation and iteratively solves the statistical equilibrium equations for the level populations, assuming a uniform-sphere geometry. Because no direct \HC--\HH\ collisional data are currently available, the \HH\ rate coefficients are obtained by scaling those for \HC--He. Since He is not a direct surrogate for \HH, the scaling must account for differences in both reduced mass and polarisability:
	\begin{equation}
		f_2=\frac{k_{\mathrm{H_2}}}{k_{\mathrm{He}}}
		=\sqrt{
			\frac{
				\alpha^{\mathrm{st}}(\mathrm{H_2})/
				\mu_{\mathrm{HC_5NH^+}\text{--}\mathrm{H_2}}
			}{
				\alpha^{\mathrm{st}}(\mathrm{He})/
				\mu_{\mathrm{HC_5NH^+}\text{--}\mathrm{He}}
			}
		}
		\label{eq:scaling}
	\end{equation}
	where $\alpha^{\mathrm{st}}(\mathrm{He})=1.384\,a_0^3$ and $\alpha^{\mathrm{st}}(\mathrm{H_2})=5.314\,a_0^3$ are the polarisabilities of He and \HH, respectively; the calculated scaling factor is $f_2=2.7$ \citep{gonzalez2024}. Because calculations involving He are computationally less demanding, He is often used as a proxy when \HH\ collision data are unavailable \citep{bop2025}. The linewidth is set to the mean of the seven transitions listed in Table~1 of \citet{marcelino2020}, approximately 0.65\,km\,s$^{-1}$.
	
\begin{figure*}
		\centering
		\includegraphics[width=\textwidth]{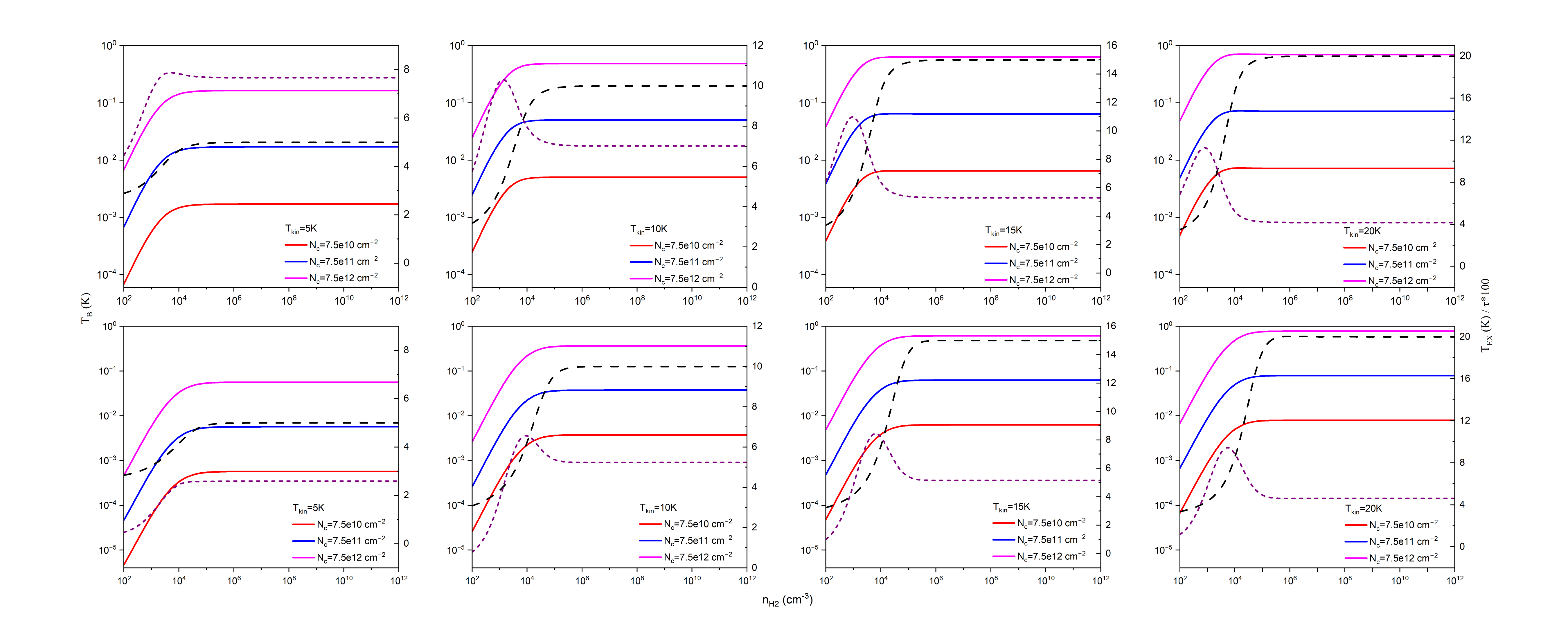}
		\caption{$T_{\mathrm{ex}}$ and $T_B$ for the $J=14\to13$ (top) and $J=18\to17$ (bottom) transitions as functions of the \HH\ volume density at three \HC\ column densities.}
		\label{fig:radex}
	\end{figure*}

	Fig.~\ref{fig:radex}(a) and Fig.~\ref{fig:radex}(b) present the non-LTE radiative transfer results for the representative transitions $J=14\to13$ (top) and $J=18\to17$ (bottom), respectively. Each transition is shown in four panels corresponding to $T_{\mathrm{kin}}=5$, 10, 15, and 20\,K. Solid lines show the brightness temperature $T_B$, long-dashed black lines the excitation temperature $T_{\mathrm{ex}}$, and short-dashed purple lines the optical depth $\tau$ multiplied by 100; the three colours correspond to the three column densities. Because $\tau<0.12$ in all models, the lines are entirely optically thin and radiative trapping is absent, and $T_B$ is approximately proportional to the column density at low densities. At the low-density ($n(\HH)\lesssim10^3$\,cm$^{-3}$), both transitions are subthermally excited ($T_{\mathrm{ex}}<T_{\mathrm{kin}}$). As $n(\HH)$ increases, collisions become increasingly dominant, and $T_{\mathrm{ex}}$ approaches $T_{\mathrm{kin}}$. $T_B$ levels off for $n(\HH)\gtrsim10^5$\,cm$^{-3}$. The variation of $\tau$ with density merely reflects differences in the populations of the upper and lower levels. Under these conditions, the plateau level of $T_B$ is determined mainly by $T_{\mathrm{ex}}$, which has approached $T_{\mathrm{kin}}$ at these densities.
	
	A comparison of the two transitions reveals a clear rotational-level effect: because $J=18\to17$ involves a higher upper-level energy and a larger energy gap, a higher density is required for it to thermalise. At the same $n(\HH)$, as long as the transitions are not fully thermalised, $T_{\mathrm{ex}}$ of $J=14\to13$ always exceeds that of $J=18\to17$; the difference is most pronounced at intermediate densities and then narrows as both lines approach $T_{\mathrm{kin}}$. For the $T_{\mathrm{kin}}=20$\,K model, at $n(\HH)=10^4$\,cm$^{-3}$, $T_{\mathrm{ex}}$ of $J=14\to13$ is already about 16.6\,K, whereas that of $J=18\to17$ is only about 8.4\,K, roughly half as high, showing that the latter thermalises significantly later. As the density rises further (e.g. to $3\times10^4$\,cm$^{-3}$), the $J=18\to17$ transition still lags behind and approaches the kinetic temperature only at about $10^5$\,cm$^{-3}$, indicating that the system is progressing towards a near-LTE state.
	
	The density at which these plateaus occur can be understood in terms of critical density ($n_{\mathrm{cr}}$). For a transition $u\to l$, the critical density is the collider density at which the rates of radiative and collisional de-excitation are equal \citep{vandertak2007}:
	\begin{equation}
		n_{\mathrm{cr}}(u\to l,T)=
		\frac{A_{u\to l}}
		{\displaystyle\sum_{l<u}k_{u\to l}(T)}
		\label{eq:ncr}
	\end{equation}
	where $u$ and $l$ denote the initial and final levels, \(A_{u\to l}\) is the Einstein A coefficient for spontaneous emission from $u$ to $l$. For the transitions considered here, we calculated the Einstein A coefficients for all \(\Delta j~=~-1\) channels within \(j \le 19\); the data are available as described in the Data availability section. Here, \(k_{u\to l}\) is the rate coefficient for collisional de-excitation from the initial level $u$ to each lower level $l$, the summation in the denominator runs over all possible de-excitation channels.
	
	Table~\ref{tab:critical} lists the values for the seven observed transitions at 5--20\,K. As the rotational quantum number increases, $A_{u\to l}$ increases rapidly with the cube of the transition frequency, and $n_{\mathrm{cr}}$ grows from $6.1\times10^3$\,cm$^{-3}$ for $13\to12$ to $1.7\times10^4$\,cm$^{-3}$ for $19\to18$; for a given transition, the $n_{\mathrm{cr}}$ varies by less than 4\% over 5--20\,K, indicating a much weaker temperature dependence than that found in neutral molecular systems (e.g. HC$_3$N--He; \citealp{wernli2007}). The typical density of TMC-1 (about $10^4$\,cm$^{-3}$) lies within this range: the low-$j$ transitions are nearly thermalised but not fully, with $T_{\mathrm{ex}}$ values slightly below $T_{\mathrm{kin}}$, and their line intensities mainly constrain the column density; the high-$j$ transitions remain subthermally excited and are sensitive to density, providing the density information. This analysis indicates the density sensitivity of each line without detailed fitting, and guides the density range of the radiative transfer grid and the joint fit. Furthermore, it demonstrates that non-LTE treatment is essential under the physical conditions of TMC-1 and that the density sensitivity is confined to the subthermal regime, where the line ratios respond to collisions.
	
	\begin{table}
		\centering
		\caption{Critical densities $n_{\mathrm{cr}}$ of the seven observed transitions at 5--20\,K.}
		\label{tab:critical}
		\small
		\begin{tabular}{lcccc}
			\hline
			\hline
			$J\to J'$ & 5\,K & 10\,K & 15\,K & 20\,K\\
			\hline
			$13\to12$ & $6.086{\times}10^{3}$ & $6.272{\times}10^{3}$ & $6.266{\times}10^{3}$ & $6.188{\times}10^{3}$\\
			$14\to13$ & $7.430{\times}10^{3}$ & $7.652{\times}10^{3}$ & $7.645{\times}10^{3}$ & $7.551{\times}10^{3}$\\
			$15\to14$ & $8.929{\times}10^{3}$ & $9.181{\times}10^{3}$ & $9.171{\times}10^{3}$ & $9.059{\times}10^{3}$\\
			$16\to15$ & $1.064{\times}10^{4}$ & $1.094{\times}10^{4}$ & $1.094{\times}10^{4}$ & $1.080{\times}10^{4}$\\
			$17\to16$ & $1.246{\times}10^{4}$ & $1.285{\times}10^{4}$ & $1.286{\times}10^{4}$ & $1.272{\times}10^{4}$\\
			$18\to17$ & $1.460{\times}10^{4}$ & $1.506{\times}10^{4}$ & $1.507{\times}10^{4}$ & $1.491{\times}10^{4}$\\
			$19\to18$ & $1.686{\times}10^{4}$ & $1.741{\times}10^{4}$ & $1.742{\times}10^{4}$ & $1.721{\times}10^{4}$\\
			\hline
		\end{tabular}
	\end{table}
	
	To constrain the observational parameters, we perform a joint fit to the seven lines observed by \citet{marcelino2020} using the $\chi^2$ statistic of \citet{lampton1976}:
	\begin{equation}
		\chi^2=\sum_i
		\left[
		\frac{I_{\mathrm{calc},i}-I_{\mathrm{obs},i}}
		{\sigma_{I,i}}
		\right]^2
		\label{eq:chi2}
	\end{equation}
	where $I_{\mathrm{calc},i}$ and $I_{\mathrm{obs},i}$ are the calculated and observed integrated line intensities, respectively, and $\sigma_{I,i}$ is the observational uncertainty.
	
\begin{figure}
		\centering
		\includegraphics[width=0.9\columnwidth]{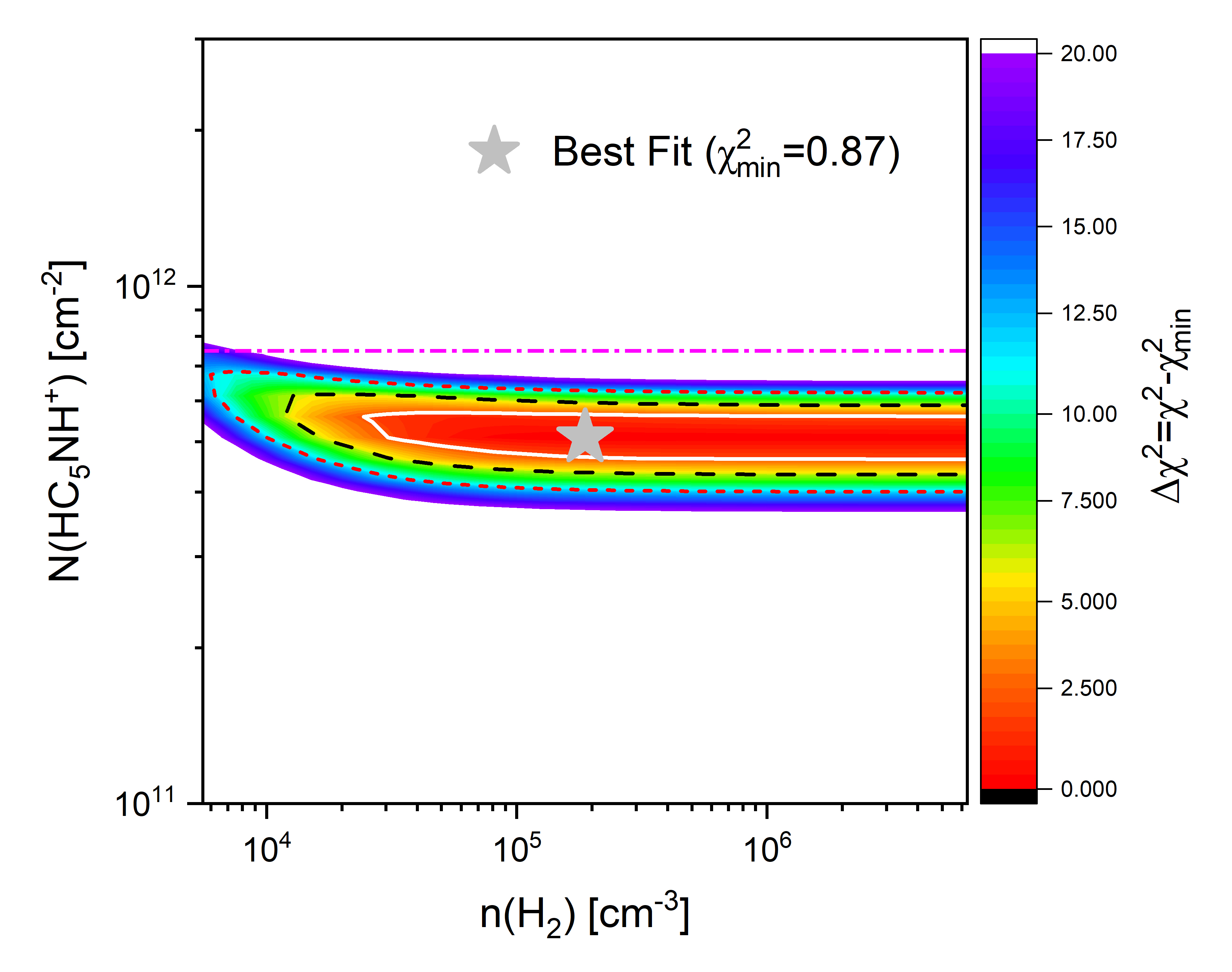}
		\caption{Distribution of $\Delta\chi^2=\chi^2-\chi^2_{\min}$ in the $n(\HH)$--$N(\HC)$ plane at $T_{\mathrm{kin}}=7.8$\,K. The best-fitting solution is marked by the silver star. The white solid, black dashed, and red short-dashed contours (from inside out) denote the $1\sigma$, $2\sigma$, and $3\sigma$ confidence level ($\Delta\chi^2=2.30$, 6.18, and 11.83, respectively).}
		\label{fig:chi2grid}
	\end{figure}

	Fig.~\ref{fig:chi2grid} shows the two-dimensional $\chi^2$ minimisation grid, where the colour scale represents $\Delta\chi^2=\chi^2-\chi^2_{\min}$ and the silver star marks the best-fitting solution; the contours, from the innermost to the outermost, correspond to $\Delta\chi^2=2.30$, 6.18, and 11.83, i.e. the $1\sigma$, $2\sigma$, and $3\sigma$ confidence regions for a two-parameter fit \citep{lampton1976}. The best fit is located at $n(\HH)=1.63\times10^5$\,cm$^{-3}$ and $N(\HC)=5.09\times10^{11}$\,cm$^{-2}$. The $\Delta\chi^2$ contours form a horizontal band: the $1\sigma$ region constrains the column density to approximately $(4.5$--$5.5)\times10^{11}$\,cm$^{-2}$ and extends towards higher densities. Because the seven lines have different $n_{\mathrm{cr}}$ and respond differently to density in the subthermal regime, the joint fit sets a lower limit on $n(\HH)$. At the best-fit density of $1.63\times10^5$\,cm$^{-3}$, which exceeds all $n_{\mathrm{cr}}$, the $T_{\mathrm{ex}}$ of the lines approaches $T_{\mathrm{kin}}$, and the contours flatten along the density direction; the resulting density degeneracy is therefore a natural consequence of the loss of diagnostic information after thermalisation, rather than a deficiency of the fitting procedure itself.
	
\begin{figure}
		\centering
		\includegraphics[width=0.9\columnwidth]{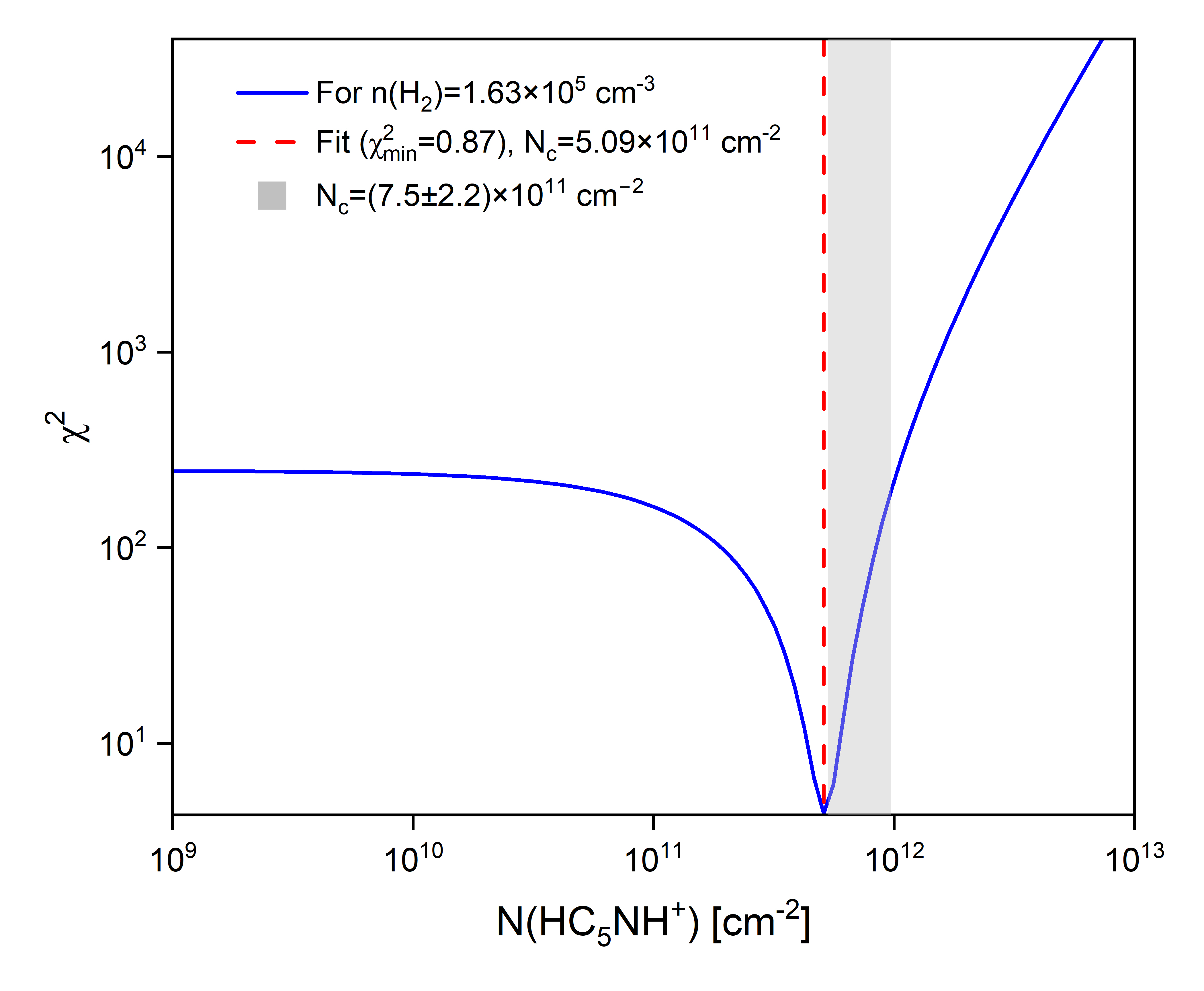}
		\caption{Variation of $\chi^2$ with the column density $N(\HC)$ at the best-fitting value of $n(\HH)$. The solid blue curve shows the fit, the dashed red line marks the best-fit column density, and the grey band shows the column density range derived from the LTE rotational diagram of \citet{marcelino2020}.}
		\label{fig:chi2profile}
	\end{figure}

	To further evaluate the goodness-of-fit and compare it with previous studies, Fig.~\ref{fig:chi2profile} displays the one-dimensional $\chi^2$ curve as a function of column density at the best-fit density of $n(\HH)=1.63\times10^5$\,cm$^{-3}$. The curve exhibits a single minimum at $N=5.09\times10^{11}$\,cm$^{-2}$, yielding a reduced $\chi^2_{\min}=0.87$. A reduced $\chi^2$ value $\le1$ is not uncommon in astrochemical analyses, typically indicating either overestimated observational uncertainties or an excellent fit to the data, as seen in other non-LTE studies (e.g. \citealp{agundez2023}). Compared to the value of $N=(7.5\pm2.2)\times10^{11}$\,cm$^{-2}$ derived from the LTE rotational diagram by \citet{marcelino2020}, our result is lower by a factor of $\sim1.5$, yet the two values are consistent within $1\sigma$. The direction and magnitude of this offset align exactly with the expected systematic difference from the LTE assumption: LTE tends to overestimate the populations of high-$j$ levels, and consequently the column density \citep{agundez2023}. The agreement in magnitude indicates that the current rate coefficients introduce no significant systematic bias in the radiative transfer modelling, thereby further validating the non-LTE treatment and the spectral line assignment.
	
	\section{Conclusions}
	\label{sec:conclusions}
	
	We have investigated the rotationally inelastic collisions of \HC\ with He and examined their implications for the excitation of \HC\ in interstellar clouds. The calculated collision dynamics exhibit a clear propensity for $|\Delta j|=2$ transitions, which is closely related to the dominant contribution of the $V_2$ component to the PES anisotropy. This highlights the important role of the interaction potential in determining the rotational transition propensities of the system.
	
	The non-LTE analysis towards TMC-1 indicates that the observed \HC\ lines are subthermally excited. While the derived column density agrees with the LTE estimate within $1\sigma$, the volume density is constrained only from below, suggesting that \HC\ is therefore a robust column-density tracer, while its density diagnostic power is limited by thermalisation.
	
	Future studies could extend this work to other homologous ions (e.g. HC$_7$NH$^{+}$) and isotopologues (e.g. HC$_5^{15}$NH$^{+}$) to investigate whether the ``even-$\Delta j$ dominance'' is a general collisional characteristic of long-chain protonated cyanopolyynes. Furthermore, the highly accurate state-to-state collisional rate coefficients presented here will provide a robust dataset to support radio observations and non-LTE analyses of HC$_{2n+1}$NH$^{+}$ and related species in molecular clouds.
	
	\section*{Data availability}
	
	The collisional rate coefficients and the associated molecular data required for the RADEX calculations will be deposited in the Zenodo repository and submitted to the LAMDA database upon acceptance for publication.
    
    \section*{Acknowledgements}

This research did not receive any specific grant from funding agencies in the public, commercial, or not-for-profit sectors.
	
	\bibliographystyle{mnras}
	\nocite{alexander1987,boys1970,manolopoulos1986,peterson2008,benabdallah2012}
	\bibliography{hc5nh}
	
	\bsp
	\label{lastpage}
\end{document}